\documentclass[11pt]{llncs}
\usepackage{fullpage}
\usepackage{graphicx}
\usepackage{url}

\usepackage{amsmath}
\usepackage{amssymb}

\newcommand{\abs}[1]{\ensuremath{\lvert #1 \rvert}}
\newcommand{\avg}[1]{\ensuremath{\overline{ #1 }}}

\newcommand{\SA}{\ensuremath{\mathsf{SA}}}
\newcommand{\DA}{\ensuremath{\mathsf{DA}}}

\newcommand{\BWT}{\ensuremath{\mathsf{BWT}}}
\newcommand{\LCP}{\ensuremath{\mathsf{LCP}}}
\newcommand{\ILCP}{\ensuremath{\mathsf{ILCP}}}

\newcommand{\find}{\textsf{find}}
\newcommand{\locate}{\textsf{locate}}

\newcommand{\rank}{\textsf{rank}}
\newcommand{\select}{\textsf{select}}
\newcommand{\doccount}{\textsf{count}}
\newcommand{\doclist}{\textsf{list}}

\newcommand{\mfind}{\ensuremath{\mathsf{find}}}
\newcommand{\mlocate}{\ensuremath{\mathsf{locate}}}
\newcommand{\mcount}{\ensuremath{\mathsf{count}}}
\newcommand{\mlist}{\ensuremath{\mathsf{list}}}
\newcommand{\mrank}{\ensuremath{\mathsf{rank}}}
\newcommand{\mselect}{\ensuremath{\mathsf{select}}}
\newcommand{\mdepth}{\ensuremath{\mathsf{depth}}}
\newcommand{\mlcp}{\ensuremath{\mathsf{lcp}}}
\newcommand{\Oh}{\ensuremath{\mathsf{O}}}
\newcommand{\oh}{\ensuremath{\mathsf{o}}}

\newcommand{\BruteD}{\textsf{Brute-D}}

\newcommand{\PDLRP}{\textsf{PDL-RP}}

\newcommand{\PDLcount}{\textsf{PDL-count}} % PDL-count
\newcommand{\SadaR}{\textsf{Sada-RR}} % Sada-R
\newcommand{\SadaPG}{\textsf{Sada-P-G}} % Sada-PG
\newcommand{\SadaPR}{\textsf{Sada-P-RR}} % Sada-PR
\newcommand{\SadaRG}{\textsf{Sada-RR-G}} % Sada-RG
\newcommand{\SadaRR}{\textsf{Sada-RR-RR}} % Sada-RR
\newcommand{\SadaG}{\textsf{Sada-grammar}} % Sada-grammar
\newcommand{\sada}{\textsf{Sada}} % sada
\newcommand{\sadaR}{\textsf{Sada-RS}} % sada_rle
\newcommand{\sadaRS}{\textsf{Sada-RS-S}} % sada_rle_sparse
\newcommand{\sadaD}{\textsf{Sada-RD}} % sada_rledv
\newcommand{\sadaDS}{\textsf{Sada-RD-S}} % sada_rledv_sparse
\newcommand{\sadaS}{\textsf{Sada-S-S}} % sada_sparse
\newcommand{\sadaSS}{\textsf{Sada-S}} % sada_sparse_simpler
\newcommand{\wt}{\textsf{ILCP}} % wt_skewed_rlebv

\newcommand{\Enwiki}{\textsf{Enwiki}}
\newcommand{\Page}{\textsf{Page}}
\newcommand{\Revision}{\textsf{Revision}}
\newcommand{\Influenza}{\textsf{Influenza}}
\newcommand{\Swissprot}{\textsf{Swissprot}}

\newcommand{\onebit}{$1$\nobreakdash-bit}
\newcommand{\zerobit}{$0$\nobreakdash-bit}

\begin{document}

\title{Document Counting in Practice\thanks{This work is funded in part by Fondecyt Project 1-140796; 
Basal Funds FB0001, Conicyt, Chile; the Jenny and Antti Wihuri Foundation, Finland; and by 
the Academy of Finland through grants 268324 and 258308.}}

\author{
Travis Gagie\inst{1}
\and
Aleksi Hartikainen\inst{1}
\and
Juha K{\"a}rkk{\"a}inen\inst{1}
\and
Gonzalo Navarro\inst{2}
\and
Simon J. Puglisi\inst{1}
\and
Jouni Sir\'en\inst{2}
}

\institute{
    Department of Computer Science,\\
    University of Helsinki, Finland\\
    \email{\{gagie,alhartik,tpkarkka,puglisi\}@cs.helsinki.fi}\\[1ex]
\and
    Center of Biotechnology and Bioengineering (CeBiB),\\
    Department of Computer Science,\\
    University of Chile, Chile\\
    \email{\{gnavarro,jsiren\}@dcc.uchile.cl}\\[1ex]
}

\date{}

\maketitle \thispagestyle{empty}
\setcounter{page}{0}

\begin{abstract}
We address the problem of counting the number of strings in a collection where
a given pattern appears, which has applications in information retrieval and
data mining. Existing solutions are in a theoretical stage. We implement
these solutions and develop some new variants, comparing them experimentally on
various datasets. Our results not only show which are the best options for 
each situation and help discard practically unappealing solutions, but 
also uncover some unexpected compressibility properties of the best data 
structures. By taking advantage of these properties, we can reduce the size of
the structures by a factor of 5--400, depending on the dataset.
\end{abstract}

\newpage

\section{Introduction}

In the classic pattern matching problem, we are given a text string $T[1,n]$ and a pattern string $P[1,m]$,
and must count or report all the positions in $T$ at which $P$ occurs.
Document retrieval problems are natural variants of this classic problem in which $T$ is composed of $d$ 
smaller strings, or {\em documents}.
The three main document retrieval problems considered to date are: {\em document counting}, 
where the task is to compute the number of documents containing $P$; {\em document listing}, where 
we must return a list of all the documents that contain $P$; and {\em top-k listing}, returning the $k$ 
documents most relevant to $P$, given some relevance measure (for example, the $k$ documents that 
contain $P$ most often). From an algorithmic point of view, these problems are interesting because
the number of occurrences of $P$ in $T$, denoted {\em occ}, may be very much larger than {\em docc}, 
the number of distinct documents in which the pattern occurs, and so tailored solutions may outperform
those based on brute-force application of classical pattern matching. 

In recent years, document retrieval problems have been the subject of intense research in both the
string algorithms and information retrieval communities (see recent surveys~\cite{HPSTV13,NavACMcs14}). The vast majority of this work has been on the latter two problems (listing and top-$k$). 
Indeed, there have been only two results on document counting~\cite{Sad07,GKNPS13}, and no investigation 
into their practicality has ever been undertaken.

However, competitive listing and top-$k$ solutions require fast algorithms for
counting. In recent work~\cite{NPS2014} it was shown that the best choice of listing and top-$k$ algorithm in 
practice strongly depends on the {\em docc}/{\em occ} ratio, and thus the ability to compute {\em docc} quickly 
may allow the efficient selection of an appropriate listing/top-$k$ algorithm at query time. Secondly, 
from an information retrieval point of view, {\em docc} (known in that community as {\em document frequency}, 
or {\em df}) is a necessary component of most ranking formulaes~\cite{ZM1998,BCC10,CMS09}, and so 
fast computation of it is desirable. Document counting is also important for data mining applications on strings 
(or {\em string mining}, see, e.g.,~\cite{FMV2008,DPT2012}), where the value {\em docc}/{\em d} for a 
given pattern is its {\em support} in the collection.

\paragraph{Our contribution.} The main results of this paper are as follows:
\begin{enumerate}
\item We provide the first implementations and experimental evaluations of the only two previous document counting
solutions by Sadakane~\cite{Sad07} and Gagie et al.~\cite{GKNPS13}. We also 
adapt to counting a previous successful structure for document listing, called
precomputed document listing (PDL)
\cite{NPS2014}. Our experiments, carried out on a wide range of data
sets, show that, while Gagie et al.'s method can use as little as a quarter of the space Sadakane's method uses,
it is always more than 10 times slower, making its use hard to justify.
Similarly, the adapted PDL is never the best choice.
\item We show that Sadakane's data structure inherits the repetitiveness present in the underlying data, which can  be exploited to reduce its space occupancy.
Surprisingly, the structure also becomes repetitive with random and near-random data, such as DNA sequences.
We show how to take advantage of this redundancy in a number of 
different ways, leading to different space-time trade-offs.
The best of these compressed representations are 5--400 times smaller than the original representation, depending on the dataset, while answering document counting queries only marginally slower, and sometimes even faster.
\end{enumerate}

The paper is organized as follows. Section~\ref{section:background} introduces some background and notation. We review Sadakane's and Gagie et al.'s methods for document counting in Section~\ref{section:algorithms}. In Section~\ref{section:new} we describe our new methods for compressing Sadakane's structure and a new variant of the PDL structure \cite{NPS2014}, adapted for document counting. Section~\ref{section:experiments} contains results and discussion from our experiments and Section~\ref{section:conclusions} concludes.

\section{Background}\label{section:background}

Let $T[1,n]$ be a concatenation of a collection of $d$ documents. We assume that each document ends with a special character $\$$ that is lexicographically smaller than any other character of the alphabet. The \emph{suffix array (SA)} of the collection is an array $\SA[1,n]$ of pointers to the suffixes of $T$ in lexicographic order. The \emph{document array (DA)} $\DA[1,n]$ is a related array, where $\DA[i]$ is the identifier of the document where suffix $T[\SA[i],n]$ begins. The \emph{suffix tree (ST)} is a versatile text index based on building a trie for the suffixes of the text, and compacting unary paths into single edges. If we list the leaves of the suffix tree in lexicographic order, we get the suffix array.

Many succinct and compressed data structures are based on \emph{bitvectors}. A bitvector is a binary sequence $B[1,n]$, with additional data structures to support \rank{} and \select. Operation $\mrank_{1}(B,i)$ counts the number of \onebit{}s in the prefix $B[1,i]$, while $\mselect_{1}(B,i)$ finds the \onebit{} of rank $i$. These operations can also be defined for \zerobit{}s, as well as on general sequences. Several different encodings are commonly used for the binary sequence. \emph{Plain} bitvectors store the sequence as-is, while \emph{entropy-compressed} bitvectors reduce its size to close to the zero-order entropy. \emph{Gap encoding} stores the distances between successive minority bits, while \emph{run-length encoding} stores the lengths of successive runs of \onebit{}s and \zerobit{}s. \emph{Grammar-compressed} bitvectors use a context-free grammar to encode the sequence.

The \emph{compressed suffix array (CSA)} \cite{GV05} and the \emph{FM-index (FMI)} \cite{FM05} are space-efficient text indexes based on the \emph{Burrows-Wheeler transform (BWT)} \cite{BW94} --- a permutation of the text originally developed for data compression. The Burrows-Wheeler transform $\BWT[1,n]$ is easily obtained from the text and the suffix array: $\BWT[i] = T[\SA[i]-1]$, with $T[0]=\$$. As the CSA and the FMI are very similar data structures, we collectively name them compressed suffix arrays (CSAs) in this paper.

We consider text indexes supporting four kinds of queries: 1) $\mfind(P)$ returns the range $[sp,ep]$, where the suffixes in $\SA[sp,ep]$ start with pattern $P$; 2) $\mlocate(P)$ returns $\SA[sp,ep]$; 3) $\mcount(P)$ returns the number of documents containing pattern $P$; 4) $\mlist(P)$ returns the identifiers of the distinct documents containing pattern $P$. For queries 2--4, we also consider variants where the parameter is the suffix array range $[sp,ep]$ or the suffix tree node $v$ corresponding to pattern $P$. CSAs support the first two queries; \find{} is relatively fast, while \locate{} can be much slower. The main time/space trade-off in a CSA, the \emph{suffix array sample period}, affects the performance of \locate{} queries. Larger sample periods result in slower and smaller indexes.

\section{Prior Methods for Document Counting}\label{section:algorithms}

In this section we review the two prior methods for document counting, one by Sadakane~\cite{Sad07} and another by Gagie et al.~\cite{GKNPS13}.

\subsection{Sadakane's method}

Sadakane \cite{Sad07} showed how to solve \doccount{} in constant time adding just $2n+\oh(n)$ bits of space. We start with the suffix tree of the text, and add new internal nodes to it to make it a binary tree. For each internal node $v$ of the binary suffix tree, with nodes $u$ and $w$ as its children, we determine the number of redundant suffixes $h(v) = \abs{\mlist(u) \cap \mlist(w)}$. This allows us to compute \doccount{} recursively: $\mcount(v) = \mcount(u) + \mcount(w) - h(v)$. By using the leaf nodes descending from $v$, $[sp,ep]$, as base cases, we can solve the recurrence:
\begin{displaymath}
\mcount(v) = \mcount(sp,ep) = (ep + 1 - sp) - \sum_{u} h(u),
\end{displaymath}
where the summation goes over the internal nodes of the subtree rooted at $v$.

We form array $H[1,n-1]$ by traversing the internal nodes in inorder and listing the $h(v)$ values. As the nodes are listed in inorder, subtrees form contiguous ranges in the array. We can therefore rewrite the solution as
\begin{displaymath}
\mcount(sp,ep) = (ep + 1 - sp) - \sum_{i=sp}^{ep-1} H[i].
\end{displaymath}
To speed up the computation, we encode the array in unary as bitvector $H'$. Each cell $H[i]$ is encoded as a \onebit, followed by $H[i]$ \zerobit{}s. We can now compute the sum by counting the number of \zerobit{}s between the \onebit{}s of ranks $sp$ and $ep$:
\begin{displaymath}
\mcount(sp,ep) = 2(ep - sp) - (\mselect_{1}(H',ep) - \mselect_{1}(H',sp)) + 1.
\end{displaymath}
As there are $n-1$ \onebit{}s and $n-d$ \zerobit{}s, bitvector $H'$ takes at most $2n+\oh(n)$ bits.

\subsection{Counting with the document array or the interleaved LCP array}
\label{sec:ilcp}

Muthukrishnan \cite{Mut02} defined, for efficiently computing $\mlist(P)$, an
array $C[1,n]$ so that $C[i]=j$ if $j$ is the rightmost position preceding 
$i$ such that $\DA[i]=\DA[j]$. He uses the property that the first occurrence
$\DA[i]$ of each document in $\DA[sp,ep]$ is the only one for which
$C[i] < sp$. This property makes $C$ useful for counting, as we only have to
determine the number of values below $sp$ in $C[sp,ep]$. This can be done in
$\Oh(\log n)$ time using a wavelet tree \cite{GGV03} on $C$, which requires 
$n\log n + \oh(n\log n)$ bits of space. Gagie et al.~\cite{GKNP13} used a more
sophisticated representation, achieving $n\log d + \oh(n\log d)+\Oh(n)$ bits of
space and query time $\Oh(\log (ep-sp+1))$ to compute $\mcount(sp,ep)$.

Both time and space are not competitive with Sadakane's method. However, a
more recent approach~\cite{GKNPS13} could be space-competitive, especially on
repetitive document collections.
Let $\mlcp(S,T)$ be the length of the \emph{longest common prefix} of
sequences $S$ and $T$. The \emph{LCP array} of $T[1,n]$ is an array
$\LCP[1,n]$, where $\LCP[i] = \mlcp(T[\SA[i-1],n], T[\SA[i],n])$. We get the
\emph{interleaved LCP array} $\ILCP[1,n]$ by building separate LCP arrays for
each of the documents, and interleaving them according to the document array.
As $\ILCP[i] < \abs{P}$ iff position $i$ contains the first occurrence of
$\DA[i]$ in $[sp,ep]$, we can solve \doccount{} by counting the number of
values less than $\abs{P}$ in $\ILCP[sp,ep]$. Just as before, this is 
efficiently done with a wavelet tree representation of $\ILCP$. The advantage 
of using $\ILCP$ is that, if the documents are similar to each other, then
$\ILCP$ will have many runs of about $d$ equal values (i.e., the same suffix
coming from all the $d$ documents), and thus it can be run-length compressed.
The wavelet tree is built only on the run heads, and $\mcount(sp,ep)$ is
computed from the run heads and the run lengths. Still this is expected to
be slower than Sadakane's method.

\section{New Techniques for Document Counting}
\label{section:new}
\subsection{Compressing Sadakane's method}

As described above, Sadakane's structures requires $2n + \oh(n)$ bits,
irrespective of the underlying data. However, even this can be a considerable
overhead on highly compressible collections, taking significantly more space
than the CSA (on top of which Sadakane's structure operates).

Fortunately, as we now show, the bitvector $H'$ used in Sadakane's method is highly compressible. There are five main ways of compressing the bitvector, with different combinations of them working better with different datasets.

\begin{enumerate}

\item Let $V_{v}$ be the set of nodes of the binary suffix tree corresponding to node $v$ of the original suffix tree. As we only need to compute $\mcount(v)$ for the nodes of the original suffix tree, the individual values of $h(u)$, $u \in V_{v}$, do not matter, as long as the sum $\sum_{u \in V_{v}} h(u)$ remains the same. We can therefore make bitvector $H'$ more compressible by setting $H[i] = \sum_{u \in V_{v}} h(u)$, where $i$ is the inorder rank of node $v$, and $H[j] = 0$ for the rest of the nodes. As there are no real drawbacks in this reordering, we will use it with all of our variants of Sadakane's method.

\item \emph{Run-length encoding} works well with versioned collections and
collections of random documents. When a pattern occurs in many documents, but no more than once in each of the documents, the corresponding subtree will be encoded as a run of \onebit{}s in bitvector $H'$.

\item When the documents in the collection have a versioned structure, we can also use \emph{grammar-based compression}. To see this, consider a substring $x$ that occurs in many documents, but at most once in each document. If each occurrence of substring $x$ is preceded by character $a$, the subtrees of the binary suffix tree corresponding to patterns $x$ and $ax$ have identical structure, and $\DA[\mfind(x)] = \DA[\mfind(ax)]$. Hence the subtrees are encoded identically in bitvector $H'$.

\item If the documents are internally repetitive but unrelated to each other,
the suffix tree has many subtrees with suffixes from just one document. We can
prune these subtrees into leaves in the binary suffix tree, using a
\emph{filter} bitvector $F[1,n-1]$ to mark the remaining nodes. Let $v$ be a
node of the binary suffix tree with inorder rank $i$. We will set $F[i] = 1$
iff $\mcount(v) > 1$. Given a range $[sp,ep-1]$ of nodes in the binary suffix tree, the corresponding subtree of the pruned tree is $[\mrank_{1}(F,sp), \mrank_{1}(F, ep-1)]$. The filtered structure consists of bitvector $H'$ for the pruned tree, and a compressed encoding of bitvector $F$.

\item We can also use filters based on array $H$ instead of $\mcount$. If $H[i] = 0$ for the most cells, we can use a \emph{sparse filter} $F_{S}[1,n-1]$, where $F_{S}[i] = 1$ iff $H[i] > 0$, and build bitvector $H'$ only for those nodes. We can also encode positions with $H[i] = 1$ separately with an \emph{$1$\nobreakdash-filter} $F_{1}[1,n-1]$, where $F_{1}[i] = 1$ iff $H[i] = 1$. With an $1$\nobreakdash-filter, we do not write \zerobit{}s in $H'$ for nodes with $H[i] = 1$, but subtract the number of \onebit{}s in $F_{1}[sp,ep-1]$ from the result of the query instead. It is also possible to use a sparse filter and an $1$\nobreakdash-filter simultaneously. In that case, we set $F_{S}[i] = 1$ iff $H[i] > 1$.

\end{enumerate}

We analyze the number of runs of \onebit{}s in bitvector $H'$ in the expected
case. Assume that our document collection consists of $d$ random documents,
each of length $m$, over an alphabet of size $\sigma$. We call string $S$ \emph{unique},
if it occurs at most once in every document. The subtree of the sparse suffix tree
corresponding to a unique string is encoded as a run of \onebit{}s in
bitvector $H'$. Therefore any set of unique strings that covers all leaves of the tree
will define an upper bound for the number of runs.

\begin{figure}[t]
\includegraphics[width=\textwidth]{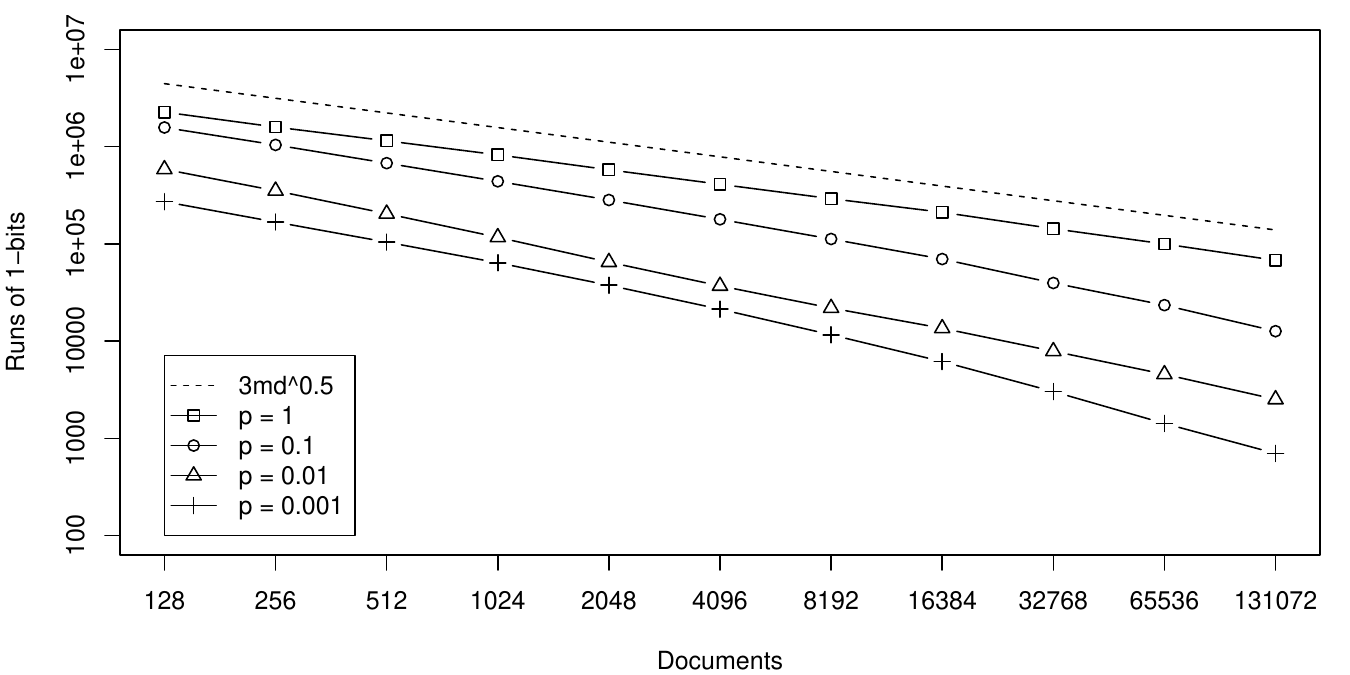}

\caption{The number of runs of $1$-bits in Sadakane's bitvector $H'$ on synthetic collections of DNA sequences ($\sigma = 4$). Each collection has been generated by taking a random sequence of length $m = 2^{7}$ to $2^{17}$, duplicating it $d = 2^{17}$ to $2^{7}$ times (making the total size of the collection $2^{24}$), and mutating the sequences with random point mutations at probability $p = 0.001$ to $1$. The dashed line represents the expected case upper bound for $p = 1$.}\label{figure:runs}
\end{figure}

Consider a random string of length $k$. The probability that the string is non-unique is
at most $dm^{2} / (2\sigma^{2k})$. Let $N(i)$ be the number of non-unique strings of length
$k_{i} = \log_{\sigma} (m \sqrt{d}) + i$. As there are $\sigma^{k_{i}}$ strings of length
$k_{i}$, the expected value of $N(i)$ is at most $m \sqrt{d} / (2 \sigma^{i})$. The expected
size of the smallest cover of unique strings is therefore at most
\begin{displaymath}
(\sigma^{k_{0}} - N(0)) + \sum_{i=1}^{\infty} (\sigma N(i-1) - N(i)) =
m \sqrt{d} + (\sigma - 1) \sum_{i=0}^{\infty} N(i) \le
\left( \frac{\sigma}{2} + 1 \right) m \sqrt{d},
\end{displaymath}
where $\sigma N(i-1) - N(i)$ is the number of strings the become unique at length $k_{i}$.
The number of runs of \onebit{}s in $H'$ is therefore sublinear in the size of the
collection ($dm$). See Figure~\ref{figure:runs} for an experimental confirmation of this analysis.

\subsection{Precomputed document counting}

\emph{Precomputed document listing (PDL)} \cite{GKNPS13} is a document listing method based on storing precomputed answers for \doclist{} queries for a carefully selected subset of suffix tree nodes. The suffix array is first covered by subtrees of the suffix tree containing a small number of leaves (e.g., no more than $256$). The roots of these subtrees are called the \emph{basic blocks}. We store the answers for \doclist{} for the basic blocks, as well as for some higher-level nodes, using grammar-based compression. The answer for $\mlist(v)$ can be found in two ways. If $v$ is below the basic blocks, the query range is short, and so the list of document identifiers can be found quickly by using the \locate{} functionality of the CSA. Otherwise we can compute $\mlist(v)$ as the union of a small number of stored answers.

A straightforward way to extend the PDL structure to support \doccount{} queries would be adding document counts to the stored answers. We now list reasons why this would not work well, and develop a structure that overcomes these problems.

\begin{enumerate}

\item Using \locate{} even for a single position is much slower than
Sadakane's method. To overcome this, we select node $v$ as a basic block if
$occ > docc$ for the node or one of its siblings, and no descendant of $v$ is
a basic block (basic blocks must form a cover of the leaves). Now if the query node is below the basic blocks, we know that $docc = occ$.

\item Computing the union of stored answers is also relatively slow. We therefore store document counts for all suffix tree nodes above the basic blocks.

\item The tree structure of PDL has been designed for speed instead of size.
When the structure stores lists of document identifiers, the tree takes only a
small fraction of the total space. When we store only document counts, this is
no longer the case. To make the tree smaller, we can use a succinct tree
representation, for example based on \emph{balanced
parentheses (BP)} \cite{MR02}.

\end{enumerate}

The \emph{precomputed document counting} structure consists of three components. Document counts are stored in an array in preorder. As most of the counts are either for the basic blocks or for nodes immediately above them, we expect them to be small compared to $d$. Therefore we use a variable-width encoding for the counts. The tree structure is stored using the balanced parenthesis representation. If there are $n'$ nodes with stored document counts, the tree takes $2n' + \oh(n')$ bits. Finally, we use a gap encoded bitvector to mark the boundaries of the basic blocks. Given a query range, we can find the corresponding range of basic blocks with \rank{} queries. Then, given the first and last basic block in the range, we can find their lowest common ancestor using standard tree operations. Finally, given the lowest common ancestor in preorder, we return the corresponding document count.

The tree only needs to support a very specific operation: namely, find the lowest common ancestor of leaves $i$ and $j$, which are guaranteed to be the first and the last leaves in the subtree. We can therefore optimize the tree for that query, making it smaller and faster than a general BP tree. Each leaf of the tree is identified by the last \onebit{} (opening parenthesis) in a run of \onebit{}s. As there are no unary internal nodes in the tree, the lowest common ancestor is either identified by the first \onebit{} in run $i$, or its closing parenthesis is the last \zerobit{} before run $j+1$, depending on which of them is deeper in the tree.

To find the lowest common ancestor, the BP bitvector needs to support two kinds of queries: \select{} for the heads of the runs of \onebit{}s, and \rank{} for the \onebit{}s. Let $\mselect(i)$ and $\mselect(j+1)$ be the starting positions of runs $i$ and $j+1$, and $\mrank(i)$ and $\mrank(j+1)$ be the ranks of the \onebit{}s at those positions. We can compute the depth of the run head $i$ as $\mdepth(i) = \mrank(i) - (\mselect(i) - \mrank(i))$, while the depth of the closing parenthesis before run $j+1$ is the same as the depth of run head $j+1$. If $\mdepth(i) \ge \mdepth(j+1)$, the lowest common ancestor of basic blocks $i$ and $j$ is node $\mrank(i)$ in preorder. Otherwise, the lowest common ancestor is node $\mrank(i) + \mdepth(j+1) - \mdepth(i)$.

\section{Experiments}\label{section:experiments}

\subsection{Implementation}

We use two fast document listing algorithms as our baseline document counting methods. \BruteD{} uses an explicit document array, sorting a copy of the query range $\DA[sp,ep]$ to count the number of distinct document identifiers. \PDLRP{} \cite{NPS2014} is a variant of precomputed document listing, using grammar-based compression to space-efficiently store the answers for \doclist{} queries for a carefully selected subset of suffix tree nodes. As the basic text index, both algorithms use RLCSA \cite{Maekinen2010}, a practical implementation of the compressed suffix array intended for repetitive datasets. The suffix array sample period was set to $32$ on non-repetitive datasets, and to $128$ on repetitive datasets.

Our implementation of precomputed document counting, \PDLcount, uses an SDSL
\cite{Gog2014b} bitvector for the tree topology, and components from the RLCSA
library for the other parts. We also used RLCSA components for several
variants of Sadakane's method. First, we have a set of basic (i.e., not
applying filtering) versions of this method, depending on how bitvector $H'$
is encoded:
\begin{description}
\item[\sada] uses a plain bitvector representation.
\item[\SadaR] uses a run-length encoded bitvector as supplied in the RLCSA
implementation. It uses $\delta$-codes to represent run lengths and packs
them into blocks of 32 bytes of encoded data. Each block stores the number
of bits and \onebit{}s up to its beginning.
\item[\sadaR] uses a run-length encoded bitvector, represented with a sparse
bitvector (like that of \sadaSS) marking the beginnings of the 0-runs and
another for the 1-runs.
\item[\sadaD] uses run-length encoding with $\delta$-codes to represent the
lengths. The bitvector is cut into blocks of 128 \onebit{}s, and three sparse
bitvectors (like in \sadaSS) are used to mark the number of bits, 1-bits, and
starting positions of block encodings. 
\item[\SadaG{}] uses grammar-compressed bitvectors \cite{NO14}.
\end{description}

There are also various versions that include filtering, and differ on how
the bitvector $F$ is represented (we only study the most promising
combinations):

\begin{description}
\item[\SadaPG] uses $\sada$ for $H'$ and a gap-encoded bitvector
for $F$. This gap-encoding is provided in the RLCSA implementation, which is
similar to that of run-length encoding but only runs of \zerobit{}s are
considered.
\item[\SadaPR] uses $\sada$ for $H'$ and a run-length encoded 
bitvector (as in \SadaR) for $F$.
\item[\SadaRG] uses $\SadaR$ for $H'$ and a gap-encoded bitvector for $F$.
\item[\SadaRR] uses $\SadaR$ for $H'$ and the same encoding for $F$.
\item[\sadaSS] uses sparse bitmaps for both $H'$ and the sparse filter $F_{S}$.
Sparse bitmaps store the lower $w$ bits of the position of each \onebit{} in an array, and use gap encoding 
in a plain bitvector for the high-order bits. Value $w$ is selected to minimize
the size (cf.~\cite{OS07}).
\item[\sadaS] is $\sadaSS$ with an additional sparse bitmap for $F_{1}$
\item[\sadaRS] uses $\sadaR$ for $H'$ and a sparse bitmap (as in \sadaSS) for
$F_{1}$.
\item[\sadaDS] uses $\sadaD$ for $H'$ and a sparse bitmap (as in \sadaSS) for
$F_{1}$.
\end{description}

Finally, \wt\ implements the technique described in Section~\ref{sec:ilcp}, 
using the same encoding as in \sadaR\ to represent the bitvectors of the wavelet tree.

The implementations were written in C++ and compiled on g++ version 4.8.1.\footnote{The implementations are available at \url{http://jltsiren.kapsi.fi/rlcsa} and \url{https://github.com/ahartik/succinct}.} Our test environment was a machine with two 2.40 GHz quad-core Xeon E5620 processors (12~MB cache each) and 96~GB memory. Only one core was used for the queries. The operating system was Ubuntu 12.04 with Linux kernel 3.2.0.

\subsection{Experimental data}\label{section:data}

We compared the performance of the document counting methods on five real datasets. Three of the datasets consist of natural language texts in XML format, while two contain biological sequences. Three of the datasets are repetitive in different ways. See Table~\ref{table:collections} for some basic statistics on the datasets.

\begin{table}[t]
\centering
\caption{Statistics for document collections. Collection size in megabytes, RLCSA size without suffix array samples in megabytes and in bits per character, number of documents, average document length, number of patterns, average number of occurrences and document occurrences, and the ratio of occurrences to document occurrences.}\label{table:collections}

\begin{tabular}{llrrrrrrrr}
\hline
\noalign{\smallskip}
Collection & \multicolumn{1}{c}{Size} & \multicolumn{1}{c}{RLCSA} & Documents & $n/d~~~$ & Patterns & $\avg{occ}~~$ & $\avg{docc}~~$ & $occ/docc$ \\
\noalign{\smallskip}
\hline
\noalign{\smallskip}
\Page      & 641 MB &   9.00 MB (0.11 bpc) &    190 & ~~~3534921 & 14286 &  2601 &     6 & 444.79 \\
\Revision  & 640 MB &   9.04 MB (0.11 bpc) &  31208 &   21490 & 14284 &  2592 &  1065 &   2.43 \\
\Enwiki    & 639 MB & ~~309.31 MB (3.87 bpc) &  44000 &   15236 & 19628 & 10316 &  2856 &   3.61 \\
\Influenza & 321 MB &  10.53 MB (0.26 bpc) & 227356 &    1480 &  1000 & ~~~59997 & ~~~44012 &   1.36 \\
\Swissprot & ~\,54 MB &  25.19 MB (3.71 bpc) & 143244 &     398 & 10000 &   160 &   121 &   1.33 \\
\noalign{\smallskip}
\hline
\end{tabular}
\end{table}

\begin{description}
\item[\Page{}] is a repetitive collection of $190$ pages with a total of $31208$ revisions from a Finnish language Wikipedia archive with full version history. The revisions of each page are concatenated to form a single document. For patterns, we downloaded a list of Finnish words from the Institute for the Languages in Finland, and chose all words of length $\ge 5$ that occur in the collection.

\item[\Revision{}] is the same as \Page, except that each revision is a separate document.

\item[\Enwiki{}] is a nonrepetitive collection of $44000$ pages from a snapshot of the English language Wikipedia. As patterns, we used search terms from an MSN query log with stop words filtered out. We generated $20000$ patterns according to term frequencies, and selected those that occur in the collection.

\item[\Influenza{}] is a repetitive collection containing the genomes of $227356$ influenza viruses. For patterns, we extracted $100000$ random substrings of length $7$, filtered out duplicates, and kept the $1000$ patterns with largest $occ/docc$ ratios.

\item[\Swissprot{}] is a nonrepetitive collection of $143244$ protein sequences used in many document retrieval papers~(e.g.~\cite{NV12}). We extracted $200000$ random substrings of length $5$, filtered out duplicates, and kept the $10000$ patterns with largest $occ/docc$ ratios.
\end{description}

\subsection{Results}

\begin{figure}[t]
\minipage{0.49\textwidth}
  \includegraphics[width=\linewidth]{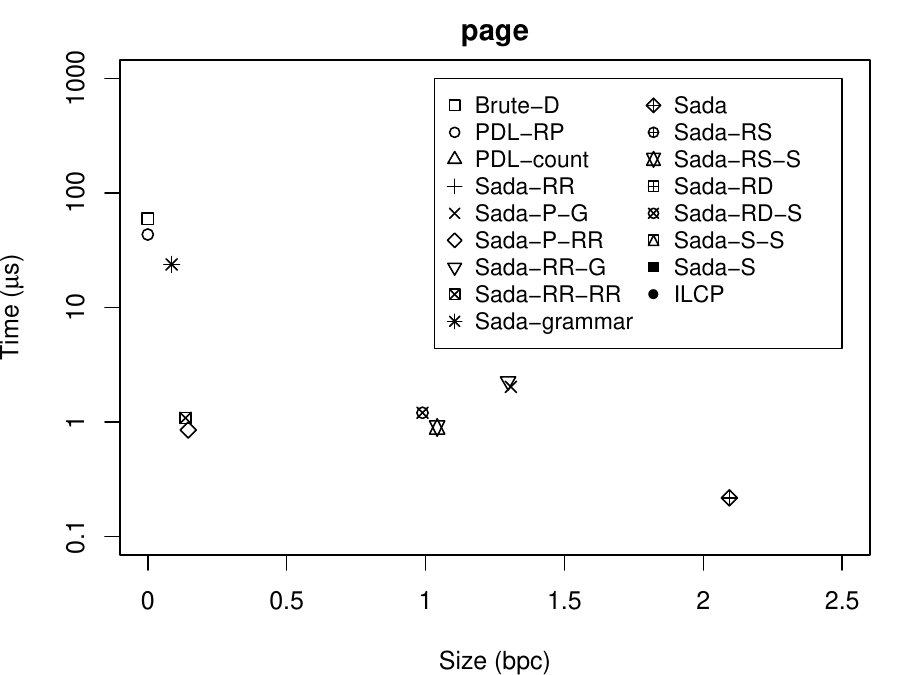}
\endminipage\hfill
\minipage{0.49\textwidth}
  \includegraphics[width=\linewidth]{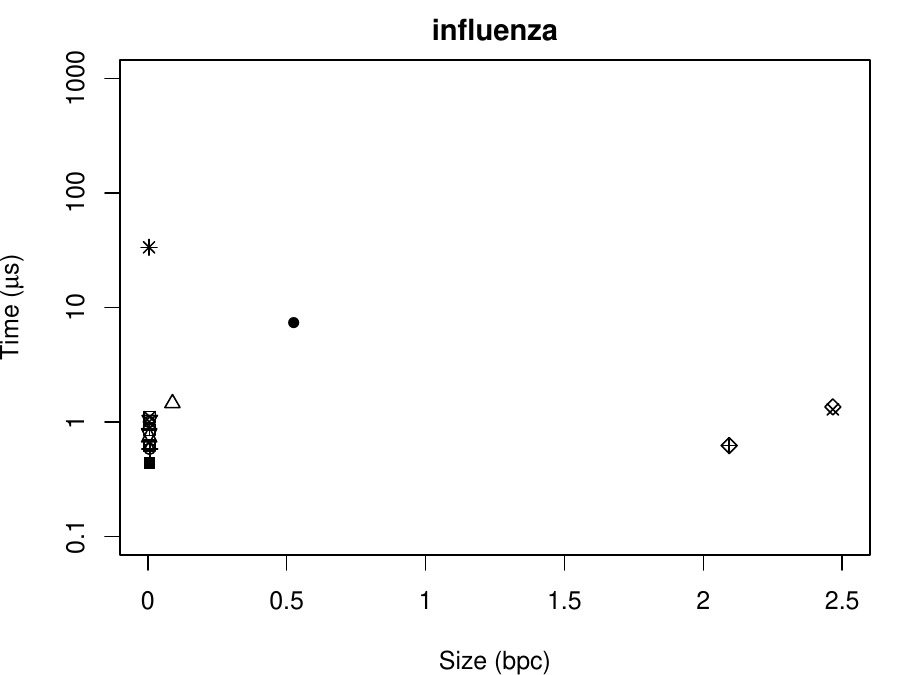}
\endminipage
\vspace{1ex}
\newline
\minipage{0.49\textwidth}
  \includegraphics[width=\linewidth]{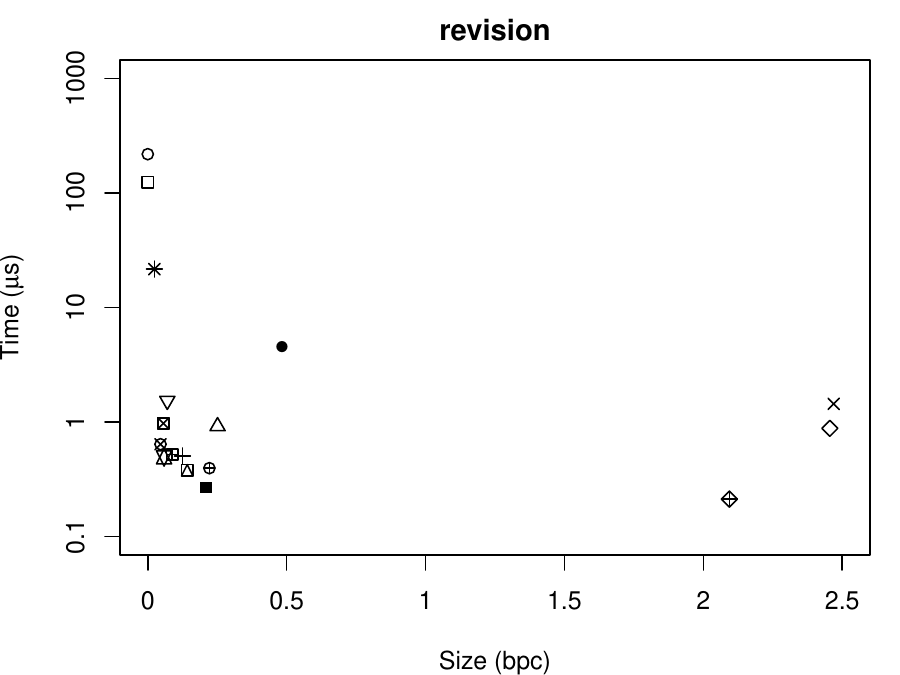}
\endminipage\hfill
\minipage{0.49\textwidth}
  \includegraphics[width=\linewidth]{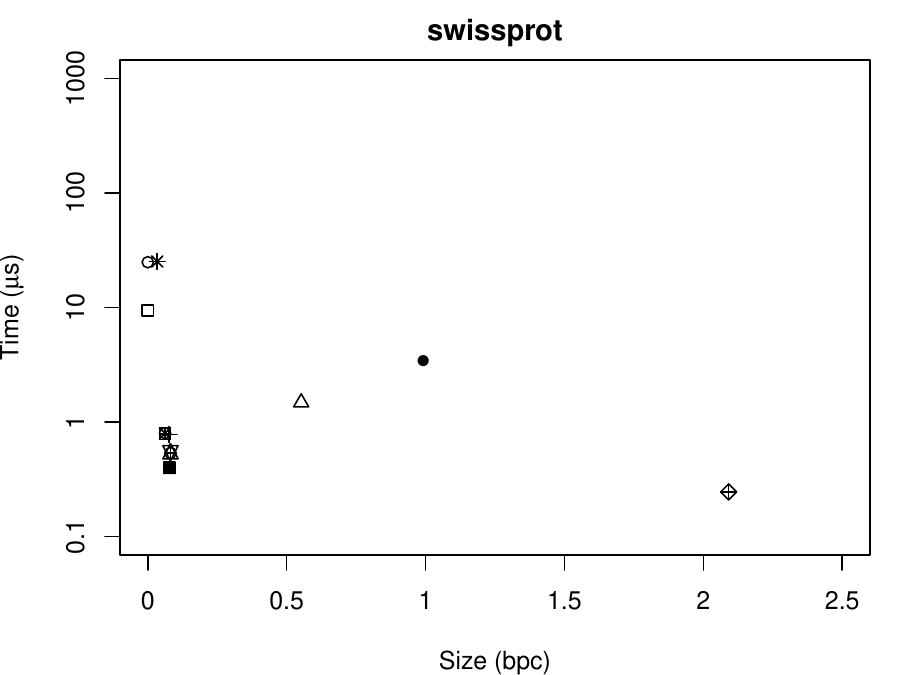}
\endminipage
\vspace{1ex}
\newline
\minipage{0.49\textwidth}
  \includegraphics[width=\linewidth]{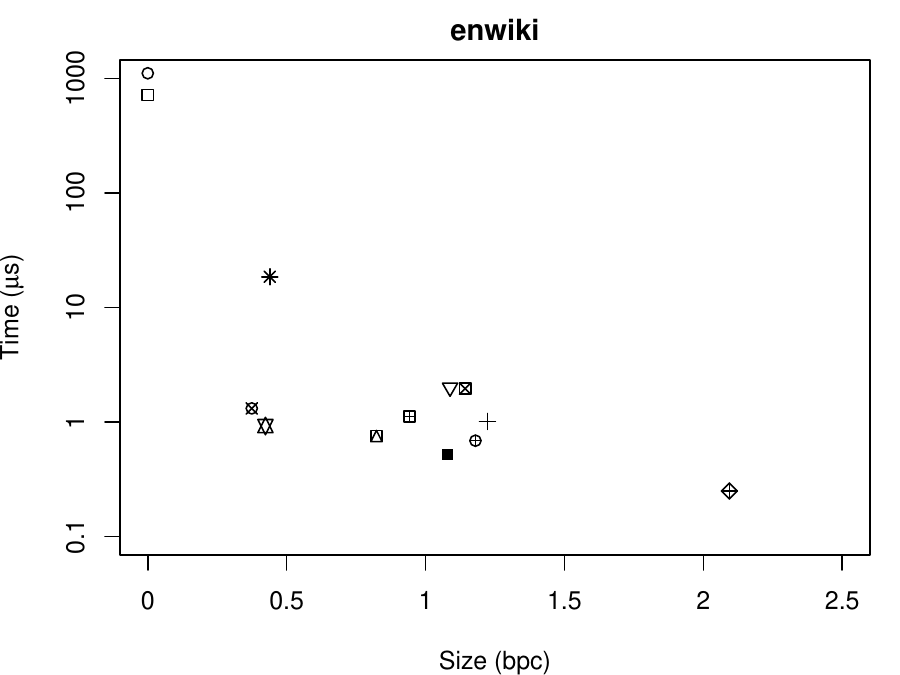}
\endminipage\hfill
\minipage{0.49\textwidth}
\endminipage
\vspace{6ex}

\caption{Document counting on different datasets. We show the average time in
microseconds required by a \doccount{} query, as a function of the size of the document counting structure in bits per character.}
\label{figure:results}
\end{figure}

The results of the experiments can be seen in Figure~\ref{figure:results}. The
time required for \find{} and the size of rest of the index (the RLCSA and possible document retrieval structures) were not included in the plots, as they are common to all solutions. As plain \sada{} was almost always the fastest method, we scaled the plots to leave out anything much larger than it. On the other hand, we set the size of the baseline document listing methods to $0$, as they are exploiting the functionality already present in the index.

On \Page{}, the filtered methods \SadaPR{} and \SadaRR{} were clearly the best choices. While plain \sada{} was much faster, it also took much more space than the rest of the index. With the exception of \SadaG{}, which was quite slow, no other method could compress the structure very well. On \Revision{}, there were many small encodings with similar performance. Among the very small encodings, \sadaRS{} was the fastest. \sadaSS{} was somewhat larger and faster. Like with \Page{}, plain \sada{} was even faster, while taking too much space to be a serious alternative.

The situation changed on the non-repetitive \Enwiki. Only \sadaDS, \sadaRS,
and \SadaG{} could compress the bitvector well below $1$~bpc, and \SadaG{} was much slower than the other two. At around $1$~bpc, \sadaSS{} was again the fastest option. Plain \sada{} required twice as much space as \sadaSS, while also being twice faster.

The other two collections, \Influenza{} and \Swissprot, contain biological (DNA and protein, respectively) sequences, and so could be considered collections of random sequences. Because such collections are easy cases for Sadakane's method, it was no surprise that many of the encodings compressed the bitvector very well. On both datasets, \sadaSS{} was the fastest small encoding, while being only marginally larger than any other encoding. As the small encodings required less than $0.01$~bpc on \Influenza{}, fitting easily in CPU cache, they were often as fast as or even faster than plain \sada.

It is interesting that different compression techniques succeed in different
collections. The only variant that is always among the smallest ones is
\SadaG, although it is much slower than most competitors.

None of the other methods could compete against a good encoding of Sadakane's method. The ILCP-based structure, \wt, was always larger and slower than compressed variants of \sada. We tried other encodings for ILCP, but they were always strictly worse than either \wt{} or plain \sada. \PDLcount{} achieved similar performance as compressed Sadakane's method on \Revision{} and \Influenza{}, but it was always somewhat larger and slower.

\section{Conclusions}\label{section:conclusions}

We investigated the time/space trade-offs in document counting data structures, implementing both known solutions and new methods. While Sadakane's method was the fastest choice, we found that it can be compressed significantly below the original $2n + \oh(n)$ bits, for a document collection of total size $n$. We achieved $5$\nobreakdash-fold compression on the natural language \Enwiki{} dataset. When the dataset was repetitive or contained random sequences, but not both, the best compressed encodings were around $20$ times smaller than the original Sadakane's structure. With both repetitive data and random sequences in the \Influenza{} collection, we achieved up to $400$\nobreakdash-fold compression. In all cases, the query times were around $1$~microsecond or less.

The high compressibility of Sadakane's structure is an unforeseen result that
emerges from our experiments. It is interesting that this compressibility owes
to very different reasons depending on the characteristics of the text
collections, but it always shows up, albeit in different degrees. 
A deeper study of the behavior of this
bitvector could uncover further compression possibilities, or lead to simple
compressibility measures on the collection that predict the ultimate size of
this representation under certain compression methods (e.g., grammar
compression, run-length compression, sparsity-based compression, etc.).

\clearpage

\bibliographystyle{plain}
\bibliography{counting}

\end{document}